\documentclass{PHYEAUTH}
\usepackage{graphicx}
\usepackage{amsmath}
\usepackage{amssymb}

\begin{document}

\begin{frontmatter}

\title{Quantum transport phenomena in disordered electron systems
with spin-orbit coupling in two dimensions and below}

\author[address1]{Yoichi Asada \thanksref{thank1}},
\author[address2]{Keith Slevin},
\author[address3]{Tomi Ohtsuki}

\address[address1]{Department of Physics, Tokyo Institute of Technology, 
2-12-1 Ookayama, Meguro-ku, Tokyo 152-8551, Japan}

\address[address2]{Department of Physics, Graduate School of Science,
Osaka University, 1-1 Machikaneyama, Toyonaka, Osaka 560-0043, Japan}

\address[address3]{Department of Physics,
Sophia University, Kioi-cho 7-1, Chiyoda-ku, Tokyo 102-8554, Japan}

\thanks[thank1]{ E-mail:asada@stat.phys.titech.ac.jp}

\begin{abstract}
Electron transport phenomena
in disordered electron systems with spin-orbit coupling
in two dimensions and below are studied numerically.
The scaling hypothesis is checked by analyzing the scaling
of the quasi-1D localization length. A logarithmic increase
of the mean conductance is also confirmed.
These support the theoretical prediction that
the two dimensional metal in systems with spin-orbit coupling
has a perfect conductivity.
Transport through a Sierpinski carpet is also reported.
\end{abstract}
\end{frontmatter}

%[main text]

%%%%%%%%%%%%%%%%%%%%%%%%%%%%%%%%%%%%%%%%%%%%%%%
\section{Introduction}
%%%%%%%%%%%%%%%%%%%%%%%%%%%%%%%%%%%%%%%%%%%%%%%

Research in spintronics
and carbon nanotubes sparked renewed
interests in the effects of spin-orbit coupling
on quantum transport \cite{zutic:04,suzuura:02}.
One of the remarkable features of non-interacting
disordered electron systems with spin-orbit coupling
is the existence of a metallic phase in two dimensions
(2D) \cite{hikami:80,ando:89}.
More exactly the 2D metallic phase appears
in systems with symplectic symmetry, i.e.,
in systems with time reversal symmetry
but without spin rotation symmetry.
Such systems are an exception to the prediction of
Abrahams {\it et al.}
that there is no metallic phase
in 2D disordered electron systems
\cite{abrahams:79}.

Recently some 2D metals have been discovered experimentally
\cite{abrahams:01}.
Although the mechanism inducing the 2D metal
might be different from spin-orbit coupling,
we think that detailed understanding
of the 2D symplectic systems gives some insights on general
transport properties of 2D metals.

A scaling theory for the symplectic systems
\cite{hikami:80,wegner:89,hikami:92}
can be described by the $\beta$
function for the conductance $g$ (in units of $e^2/h$)
\begin{equation}
  \beta(\ln g)=\frac{\mathrm{d}\ln g}{\mathrm{d}\ln L},
  \label{eq:betag}
\end{equation}
where $L$ is the linear size of a system.
(The conductance is actually a distributed quantity.
See Refs.~\cite{shapiro:87,slevin:01,slevin:03}
for discussions on the scaling hypothesis
of the conductance.)
Because of the so-called anti-localization effect,
the $\beta$ function shows non-monotonic behavior
in the symplectic systems.
The asymptotic behavior in the strongly metallic
region in 2D is conjectured to be
\begin{equation}
 \beta(\ln g)=\frac{1}{\pi g} \hspace{5mm}
 \left(g\gg 1\right).
 \label{eq:betaasymptotic}
\end{equation}
This indicates that there is a metallic phase in 2D.

In the scaling theory for the symplectic systems in 2D,
however, there is a long standing problem
for more than two decades
\cite{kawabata:88,kawarabayashi:96}:
the logarithmic divergence of the conductance
in the 2D metallic phase
in the limit $L\rightarrow \infty$
indicated by Eq.~(\ref{eq:betaasymptotic}).
Since the conductivity and the conductance are the same quantity
in 2D, this predicts a perfect conductivity, $\sigma=\infty$,
in spite of the system being disordered.
This is in contrast to 3D disordered metals,
in which the conductivity converges to a finite value.

According to the renormalization group theory of continuous phase
transitions, the scaling hypothesis is valid when the correlation
length is much larger than all other microscopic lengths.
The scaling argument which deduced (\ref{eq:betaasymptotic}),
however, is based on the weak localization  theory
in which disorder is supposed to be weak.
In such a weakly disordered metallic phase this condition
on the correlation length may not be satisfied.
Therefore, a numerical check of the scaling hypothesis
in the 2D metallic region is desirable.

In this paper, we report a numerical check of
the scaling hypothesis in the 2D metallic region.
We also report numerical calculation of the
Landauer conductance in 2D,
which is a direct check of the prediction of a
perfect conductivity.
Last, numerical simulations of transport through a Sierpinski
carpet are reported.

%%%%%%%%%%%%%%%%%%%%%%%%%%%%%%%%%%%%%%%%%%%%%%%
\section{The SU(2) model}
%%%%%%%%%%%%%%%%%%%%%%%%%%%%%%%%%%%%%%%%%%%%%%%
Among the models with symplectic symmetry,
we employ the SU(2) model \cite{asada:02,asada:04},
\begin{equation}
 H=\sum_{i,\sigma}\epsilon_i c_{i\sigma}^{\dagger} c_{i\sigma}
- \sum_{\langle i, j\rangle,\sigma,\sigma'}
R(i,j)_{\sigma\sigma'} c_{i\sigma}^{\dagger} c_{j\sigma'}.
\end{equation}
The random potential $\epsilon_i$
is distributed with box distribution
in the range $[-W/2,W/2]$.
Random spin-orbit coupling is included in the nearest neighbor
hoping term.
We parameterize the hopping matrix as
\begin{eqnarray}
 R(i,j)=\left(
 \begin{array}{cc}
  \mathrm{e}^{\mathrm{i}\alpha_{ij}}\cos\beta_{ij}
  & \mathrm{e}^{\mathrm{i}\gamma_{ij}} \sin\beta_{ij} \\
  -\mathrm{e}^{-\mathrm{i}\gamma_{ij}}\sin\beta_{ij}
  & \mathrm{e}^{-\mathrm{i}\alpha_{ij}}\cos\beta_{ij}
 \end{array}
 \right),
\end{eqnarray}
and we distribute
$\alpha_{ij}$ and $\gamma_{ij}$ with uniform probability
in the range $\left[0,2\pi\right)$, and
$\beta_{ij}$ according to the probability density
$p(\beta) {\mathrm d} \beta =  \sin (2\beta) {\mathrm d} \beta$
in the range $\left[0,\pi/2 \right]$.
In the actual simulations $R(i,j)$'s in one direction are set
to the unit matrix with the aid of the local SU(2) gauge
transformation.

The SU(2) model has an advantage that corrections to scaling
arising from irrelevant variables are smaller than other models
\cite{asada:02,asada:04},
so it is quite a useful model
when we study universal aspects of
Anderson localization and the Anderson transition.

%%%%%%%%%%%%%%%%%%%%%%%%%%%%%%%%%%%%%%%%%%%%%%%
\section{Numerical check of the scaling hypothesis in the 2D metallic phase}
%%%%%%%%%%%%%%%%%%%%%%%%%%%%%%%%%%%%%%%%%%%%%%%

We analyze the quasi-1D localization length $\lambda$
on a quasi-1D strip with width $L$
\cite{pichard:81,mackinnon:83}.
Periodic boundary conditions are imposed in the transverse
direction.
The scaling hypothesis implies that
the dimensionless quantity $\Lambda=\lambda/L$ obeys
\begin{equation}
 \Lambda=F_{\pm}\left(\frac{L}{\xi}\right),
\end{equation}
where $\xi$ is the 2D correlation (localization) length
which depends on $W$ and the Fermi energy $E$.
The subscript $\pm$ indicates the metallic
and insulating phases.

We have calculated $\Lambda$
for sizes $L=[16,128]$
with an accuracy from $0.3\%$ to $1.0\%$.
The ensemble transfer matrix method has been used \cite{slevin:04}.
In addition to the data at $E=1$ analyzed in Ref.~\cite{asada:04},
we have also accumulated data at $E=2$ in $W=[0.0,4.5]$.
From the numerical data
the correlation length $\xi$ at each pair of $(E,W)$
and the scaling function $F_+(L/\xi)$ are
estimated using numerical fit
described in Ref.~\cite{asada:04}.
The quality of the fit is assessed by
the goodness of fit probability $Q$.
To eliminate the ambiguity of the absolute scale of $\xi$,
we set $\xi=10$ at $(E,W)=(1,3)$.

\begin{figure}[tb]
\includegraphics[width=\linewidth]{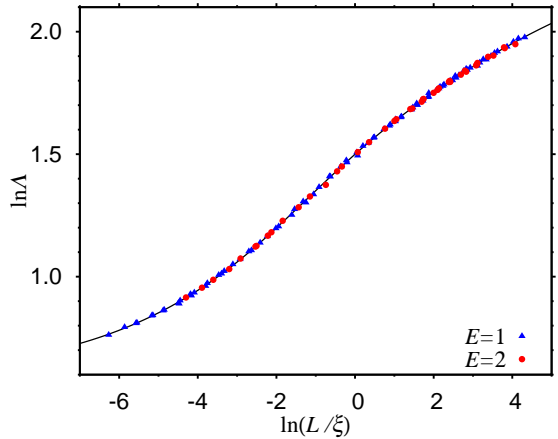}
\caption{Single parameter scaling in the metallic region
in 2D.
The goodness of fit $Q$ is $0.3$.
}
\label{fig:metalsps1}
\end{figure}

\begin{figure}[tb]
\includegraphics[width=\linewidth]{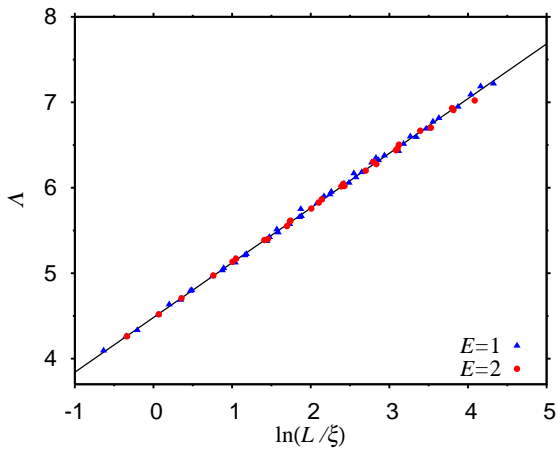}
\caption{Single parameter scaling in the strongly metallic region
in 2D. The goodness of fit $Q$ is $0.7$.
}
\label{fig:metalsps2}
\end{figure}

Figure~\ref{fig:metalsps1}
demonstrates the single parameter scaling in the metallic region.
All data fall on a common scaling curve when $\Lambda$ is
plotted as a function of $L/\xi$,
indicating the validity of the scaling hypothesis.
In a strongly metallic  region ($\Lambda\gg 1$),
the data are well fitted to
\begin{equation}
 \Lambda=b+c\ln (L/\xi) \hspace{5mm}(\Lambda \gg 1),
 \label{eq:lambdaln}
\end{equation}
as shown in Fig.~\ref{fig:metalsps2}.

In Ref.~\cite{asada:04}, the $\beta$ function for $\Lambda$
from the metallic to the insulating limits was estimated.
The estimate has not changed significantly when we add data
for $E=2$.

%%%%%%%%%%%%%%%%%%%%%%%%%%%%%%%%%%%%%%%%%%%%%%%
\section{Logarithmic increase of the conductance
in the 2D metallic phase}
%%%%%%%%%%%%%%%%%%%%%%%%%%%%%%%%%%%%%%%%%%%%%%%

In last section and in Ref.~\cite{asada:04},
the scaling hypothesis has been checked numerically.
This supports the scaling theory for the 2D symplectic systems.
Furthermore, the logarithmic increase of $\Lambda$
in the strongly metallic region 
gives the impression that the  conductance
may also increase logarithmically.
In this section, we study the size dependence
of the mean conductance directly.

We now attach two perfect leads to the square sample
with linear size $L$
and calculate the two terminal Landauer conductance $g$
(in units of $e^2/h$) with the transfer
matrix method \cite{pendry:92}.
A fixed boundary condition is imposed in the direction
transverse to the current.
The hopping matrices in the transverse direction
are set to the unit matrix,
and the Fermi energy $E=1$.
We have accumulated 10000 samples for
$L=[64,256]$ and 5000 samples for $L=384$.

\begin{figure}[tb]
\includegraphics[width=\linewidth]{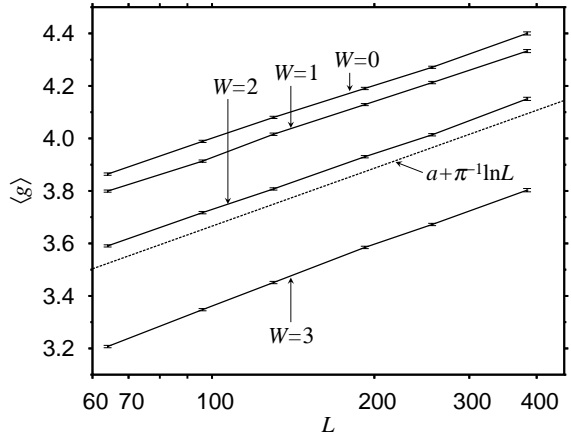}
\caption{The mean conductance
$\langle g \rangle$ vs. system size $L$ in the 2D metallic region.
The solid lines are a guide to the eye only.
For a reference,
$\langle g \rangle =a + \pi^{-1}\ln L$ is indicated.
}
\label{fig:2dmeang}
\end{figure}

As shown in Fig.~\ref{fig:2dmeang},
the mean conductance $\langle g \rangle$ in the
strongly metallic region does show
the logarithmic increase
\begin{equation}
 \langle g \rangle \approx a + \pi^{-1}\ln L
 \hspace{5mm}(\langle g \rangle \gg 1),
\end{equation}
in agreement with the prediction of the scaling theory
\cite{hikami:80}.
The prefactor of the logarithmic term,
which is an important parameter for the scaling
in the strongly metallic region,
is also consistent with it.
Note that a perfect conductivity
does not mean perfect transmission.
The transmission probability per channel rather goes to
zero as $L$ increases, so most of incoming electrons are
reflected.

It is left for future to understand physics
behind this possible perfect conductivity in 2D.

%%%%%%%%%%%%%%%%%%%%%%%%%%%%%%%%%%%%%%%%%%%%%%%
\section{Transport through a Sierpinski carpet}
%%%%%%%%%%%%%%%%%%%%%%%%%%%%%%%%%%%%%%%%%%%%%%%

Recently we have obtained numerical results
which suggest that an Anderson transition
occurs even below 2D in the presence of
spin-orbit coupling \cite{asada:below2d}.
This is based on numerical simulations
of electrons on a Sierpinski carpet SC($5,1,k$),
where $k$ is the generation number.
In the limit $k\rightarrow\infty$,
SC($5,1,k$) becomes a true fractal whose
spectral dimension is
$d_{\mathrm s}=1.940\pm 0.009$ \cite{fujiwara:95}.

Here we study the size dependence
of the Landauer conductance through
the Sierpinski carpet in a delocalized phase.
We have attached two perfect leads to SC($5,1,k$)
and have calculated the conductance
with the recursive Green's function method \cite{ando:91}.
A fixed boundary condition is imposed in the
transverse direction.
The hopping matrices in the transverse direction are
set to the unit matrix.
We set the Fermi energy $E=0.9$ and the disorder $W=0,2$
where relatively
large mean conductance is obtained.
The number of samples is 10000 for $k=1,2,3$ and
500 for $k=4$.

\begin{figure}[tb]
\includegraphics[width=\linewidth]{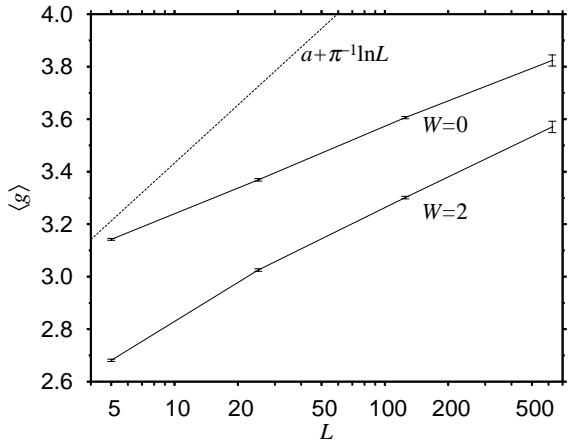}
\caption{The mean conductance
$\langle g \rangle$ vs. system size $L=5^k$ for
the Sierpinski carpet SC($5,1,k$).
The solid lines are a guide to the eye only.
The dashed line indicates the logarithmic increase in 2D.
}
\label{fig:SC51meang}
\end{figure}

Figure~\ref{fig:SC51meang} shows the mean conductance
$\langle g \rangle $ as a function of the linear size $L=5^k$.
The mean conductance increases with $L$,
possibly indicating a delocalized phase.
It also indicates that the increase
is slower than that in 2D.
It is an open problem whether
the conductance in the possible delocalized phase of
the Sierpinski carpet diverges.

\begin{ack}
%{\bf Acknowledgements}:
Yoichi Asada would like to thank Professor Tsuneya Ando
for valuable discussions.
He is supported by Research Fellowships of the Japan
Society for the Promotion of Science for Young Scientists.
\end{ack}


\begin{thebibliography}{9}

\bibitem{zutic:04} I. \v{Z}uti\'{c}, J. Fabian, and S. Das Sarma,
Rev. Mod. Phys. \textbf{76}, 323 (2004).

\bibitem{suzuura:02} H. Suzuura and T. Ando,
Phys. Rev. Lett. {\bf 89}, 266603 (2002).


\bibitem{hikami:80} S. Hikami, A. I. Larkin, and Y. Nagaoka,
Prog. Theor. Phys. {\bf 63}, 707 (1980).

\bibitem{ando:89} T. Ando, Phys. Rev. B {\bf 40}, 5325 (1989).


\bibitem{abrahams:79}E. Abrahams, P. W. Anderson,
D. C. Licciardello, and T. V. Ramakrishnan,
Phys. Rev. Lett. {\bf 42}, 673 (1979).

\bibitem{abrahams:01}
E. Abrahams, S. V. Kravchenko, and M. P. Sarachik,
Rev. Mod. Phys. {\bf 73}, 251 (2001).

\bibitem{wegner:89}  F. Wegner,
Nucl. Phys. {\bf B316}, 663 (1989).

\bibitem{hikami:92} S. Hikami,
Prog. Theor. Phys. Suppl. {\bf 107}, 213 (1992).

\bibitem{shapiro:87} B. Shapiro, Phil. Mag. B {\bf 56},
1031 (1987).

\bibitem{slevin:01} K. Slevin, P. Marko\v{s}, and T. Ohtsuki,
Phys. Rev. Lett. {\bf 86}, 3594 (2001).

\bibitem{slevin:03} K. Slevin, P. Marko\v{s}, and T. Ohtsuki,
Phys. Rev. B {\bf 67}, 155106 (2003).

\bibitem{kawabata:88} A. Kawabata,
J. Phys. Soc. Jpn. {\bf 57}, 1717 (1988).

\bibitem{kawarabayashi:96} T. Kawarabayashi and T. Ohtsuki,
Phys. Rev. B {\bf 53}, 6975 (1996).


\bibitem{asada:02}Y. Asada, K. Slevin, and T. Ohtsuki,
Phys. Rev. Lett. {\bf 89}, 256601 (2002).

\bibitem{asada:04}Y. Asada, K. Slevin, and T. Ohtsuki,
Phys. Rev. B {\bf 70}, 035115 (2004).

\bibitem{pichard:81} J-L. Pichard and G. Sarma,
J. Phys. C {\bf 14}, L127 (1981).

\bibitem{mackinnon:83} A. MacKinnon and B. Kramer,
Z. Phys. B {\bf 53}, 1 (1983).

\bibitem{slevin:04}K. Slevin, Y. Asada, and L. I. Deych,
Phys. Rev. B {\bf 70}, 054201 (2004).


\bibitem{pendry:92}J. B. Pendry, A. MacKinnon, and P. J. Roberts,
Proc. R. Soc. London, Ser. A
{\bf 437}, 67 (1992).

\bibitem{asada:below2d}Y. Asada, K. Slevin, and T. Ohtsuki,
arXiv cond-mat/0508315.

\bibitem{fujiwara:95} S. Fujiwara and F. Yonezawa,
Phys. Rev. E {\bf 51}, 2277 (1995).

\bibitem{ando:91} T. Ando,
Phys. Rev. B {\bf 44}, 8017 (1991).

\end{thebibliography}
\end{document}